\documentclass[symmetry,article,submit,moreauthors,pdflatex,10pt,a4paper]{mdpi} 
\preto{\abstractkeywords}{\nolinenumbers}
\firstpage{1} 
\pubvolume{xx}
\pubyear{2019}
\copyrightyear{2019}
\history{Received: date; Accepted: date; Published: date}

\def\al{\alpha}
\def\be{\beta}
\def\ga{\gamma}
\def\de{\delta}

\def\th{\theta}

\def\la{\lambda}

\def\si{\sigma}

\def\ch{\chi}
\def\ps{\psi}
\def\om{\omega}

\def\De{\Delta}

\def\Si{\Sigma}

\def\cH{{\cal H}}
\def\cl{{\cal L}}

\def\fr#1#2{{{#1}\over{#2}}}
\def\frac#1#2{{\textstyle{{#1}\over{#2}}}}
\def\half{{\textstyle{1\over 2}}}
\def\ol{\overline}
\def\prt{\partial}

\def\lsim{\mathrel{\rlap{\lower4pt\hbox{\hskip1pt$\sim$}}
    \raise1pt\hbox{$<$}}}
\def\gsim{\mathrel{\rlap{\lower4pt\hbox{\hskip1pt$\sim$}}
    \raise1pt\hbox{$>$}}}

\def\vev#1{\langle {#1}\rangle}

\def\nn{\nonumber}

\def\psb{\ol\ps{}}

\def\mbf#1{\boldsymbol #1}

\def\Q{\mathcal Q}

\def\pvec{\mbf p}
\def\gavec{\mbf\ga}

\def\Qhat{\widehat\Q}

\def\Z{Z}

\def\cmtemplate#1#2#3#4{{#1}^{#3}_{#4}}

\def\ctemplate#1#2#3#4{{#1}^{(#2)#3}_{#4}}

\def\bcf#1#2{\ctemplate{b}{#1}{#2}{F}}

\def\cmtemplate#1#2#3#4{{#1}^{#3}_{#4}}

\def\bcmw#1#2#3{\cmtemplate{b}{#1}{#2}{{#3}}}

\def\dcmw#1#2#3{\cmtemplate{d}{#1}{#2}{{#3}}}

\def\gcmw#1#2#3{\cmtemplate{g}{#1}{#2}{{#3}}}
\def\Hcmw#1#2#3{\cmtemplate{H}{#1}{#2}{{#3}}}

\def\ctemplate#1#2#3#4{{#1}^{(#2)#3}_{#4}}

\def\bcw#1#2#3{\ctemplate{b}{#1}{#2}{{#3}}}

\def\dcw#1#2#3{\ctemplate{d}{#1}{#2}{{#3}}}

\def\gcw#1#2#3{\ctemplate{g}{#1}{#2}{{#3}}}
\def\Hcw#1#2#3{\ctemplate{H}{#1}{#2}{{#3}}}

\def\bcfw#1#2#3{\ctemplate{b}{#1}{#2}{F{,#3}}}

\def\dcfw#1#2#3{\ctemplate{d}{#1}{#2}{F{,#3}}}

\def\gcfw#1#2#3{\ctemplate{g}{#1}{#2}{F{,#3}}}
\def\Hcfw#1#2#3{\ctemplate{H}{#1}{#2}{F{,#3}}}

\def\ab{{\al\be}}

\def\mab{{\mu\al\be}}

\def\m{m_\ps}
\def\mw{m_w}
\def\Z{\si}

\def\btw#1#2{{\widetilde b}_{#1}^{#2}}
\def\bftw#1#2{{\widetilde b}_{F,#1}^{#2}}

\def\btws#1#2{{\widetilde b}_{#1}^{*#2}}
\def\bftws#1#2{{\widetilde b}_{F,#1}^{*#2}}

\newcommand*\labeleeq[1]{ \label{#1} \end{equation} } 
\newcommand*\labeleea[1]{ \label{#1} \end{eqnarray} }

\newcommand{\beq}{\begin{equation}}
\newcommand{\eeq}{\end{equation}}
\newcommand{\bea}{\begin{eqnarray}}
\newcommand{\eea}{\end{eqnarray}}
\newcommand{\rf}[1]{(\ref{#1})}
\newcommand{\Ref}[1]{Ref.~\cite{#1}}
\newcommand{\eqs}[1]{Eqs.~(\ref{#1})}
\newcommand{\eq}[1]{Eq.~(\ref{#1})}

\Title{Lorentz and CPT Tests using Penning Traps}

\Author{Yunhua Ding}

\AuthorNames{Yunhua Ding}

\address[1]{
Department of Physics,
Gettysburg College, Gettysburg, PA 17325; yding@gettysburg.edu} 

\abstract{
The theoretical prospects for quantum electrodynamics 
with Lorentz-violating operators of mass dimensions up to six are revisited in this work.
The dominant effects due to Lorentz and CPT violation are studied
in measurements of magnetic moments of particles confined in Penning traps.
Using recently reported experimental results,
new coefficients for Lorentz violation are constrained
and existing bounds of various coefficients are improved.
}


\begin{document}

\section{Introduction}
\label{Introduction}

The recent measurements of the proton and antiproton magnetic moments 
have reached record sensitivities of ppb levels 
by confining the particles in electromagnetic fields using a Penning trap
\cite{17sc, 17sm}.
For the electron magnetic moment,
a similar Penning-trap experiment has also been carried out
in an impressive precision of ppt level
\cite{ha11}.
Experiments measuring the positron magnetic moment 
are currently under development to aim for a comparable precision as that of the electron
\cite{15fo, 19gf}.
These highly precise measurements in Penning-trap experiments 
offer a great way to study fundamental symmetries, 
including Lorentz and CPT invariances,
the foundation of Einstein's theory of relativity.
It has been shown that tiny deviations from relativity could naturally emerge 
in a fundamental theory unifying gravity with quantum physics 
at the Plank scale $M_P\sim 10^{19}$~GeV,
such as strings
\cite{89ks,91kp}.
In recent years,
many high-precision tests of relativity in various subfields of physics
have been performed to search for Lorentz- and CPT-violating signals
\cite{dt},
including the spectroscopic studies of particles confined in Penning traps.

Any tiny violation effects arising from a large unknown energy scale 
are well described by effective field theory. 
The comprehensive framework describing Lorentz violation 
in the context of effective field theory 
is given by the Standard-Model Extension (SME)
\cite{ck,ck1,akgrav},
which is constructed by adding all possible Lorentz-violating terms 
into the action of General Relativity and the Standard Model.
Each of the terms is formed from a coordinate-independent contraction
of a Lorentz-violating operator with the corresponding controlling coefficient.
In the context of effective field theory,
CPT violation is accompanied by the breaking of Lorentz symmetry
\cite{ck,ck1,owg},
so the SME also describes general CPT-violating effects.
The SME provides a general framework to study possible effects 
due to Lorentz and CPT violation and
the parameters in any specific model characterizing these violations
can be matched to a suitable subset of the SME coefficients.

The minimal SME contains Lorentz-violating operators of mass dimensions up to four,
which is power-counting renormalizable in Minkowski spacetime.
Lorentz-violating operators of larger mass dimensions can be viewed
as corrections at higher orders to the low-energy limit.
Study of the nonminimal SME is of interest 
in many different contexts of physics,
such as the causality and stability 
\cite{akrl,causality},
the pseudo-Riemann-Finsler geometry 
\cite{finsler, finsler1,finsler2, 19ek},
the mixing of Lorentz-violating operators of different mass dimensions 
\cite{clp14},
Lorentz-violating models in supersymmetry
\cite{susy},
and noncommutative Lorentz-violating quantum electrodynamics 
\cite{chklo,ncqft,ncqft2}.

In a Penning-trap experiment,
the measurable effects due to Lorentz and CPT violation 
given by the minimal SME include 
the charge-to-mass ratio and the magnetic moment 
of the confined particle 
\cite{bkr97,bkr98},
through the changes in the anomaly and cyclotron frequencies.
The published work on studying the minimal-SME effects involves
comparison of the anomaly frequencies of the electron and positron 
\cite{de99},
sidereal-variation analysis of the electron anomaly frequency
\cite{mi99, 07ha},
and measurements of cyclotron frequencies
of the H$^-$ ion relative to that of the antiproton
\cite{ga99,ul15}.

In the nonminimal SME,
additional Lorentz- and CPT-violating effects beyond the minimal SME can be generated
by the interaction between the confined particle and the electromagnetic fields in the trap.
The general theory of quantum electrodynamics with Lorentz- and CPT-violating operators 
of mass dimensions up to six has been constructed in \Ref{16dk}.
Recently,
this treatment was generalized to include operators of arbitrary mass dimension
using gauge field theory
\cite{19kl}.
In this work we focus on further studies of Lorentz- and CPT-violating effects 
in the nonminimal sector of the SME by using the recent Penning-trap results,
which include the sidereal-variation analysis of the anomaly frequencies for electrons 
\cite{07ha}
and the comparison of magnetic moments for both protons and antiprotons 
\cite{17sm, 17sc},
to obtain new and improved constraints on the SME coefficients.
The results from this work are complementary to 
existing studies of Penning-trap experiments
\cite{16dk},
the muon anomalous magnetic moment
\cite{muon,gkv14},
clock-frequency comparison
\cite{kv18},
and spectroscopy of hydrogen, antihydrogen, and other related systems
\cite{kv15}. 

This work is organized as follows.
In Section~\ref{theory},
we review the theory of quantum electrodynamics 
with Lorentz- and CPT-violating operators of mass dimensions up to six.
We use perturbation theory to obtain the dominant shifts arising from Lorentz violation
to the energy levels of the trapped fermion,
and then derive the contributions to the cyclotron and anomaly frequencies.
The discussion of coordinate transformation
is given at the end of this section.
We next turn in Section~\ref{experiment}
the experimental applications related to Penning traps 
and use the reported results to extract new limits on various SME coefficients,
including some that were not constrained in the literature.
The constraints on the SME coefficients
obtained in this work are summarized in Section~\ref{sensitivity}.

\section{Theory}
\label{theory}

The theoretical prospects of Lorentz- and CPT-violating
quantum electrodynamics in Penning-trap experiments
have been studied in ~\Ref{16dk}.
In this section we review the main results.

For a Dirac fermion field $\ps$ of charge $q$ and mass $\m$
confined in an external electromagnetic field specified by potential $A_\mu$,
the conventional gauge-invariant Lagrange density $\cl_0$ takes the form
\beq
\cl_0 =
\half \psb (\ga^\mu i D_\mu - \m ) \ps + {\rm h.c.} , 
\label{fermlag0}
\eeq
where the covariant derivative is given by the minimal coupling
$iD_\mu = (i \prt_\mu - q A_\mu)$
and h.c. means hermitian conjugate.
The general Lorentz-violating Lagrange density that preserves U(1) gauge invariance
for the Dirac fermion field $\ps$ can be constructed by adding contraction terms
of Lorentz-violating operators with the corresponding SME coefficients
\cite{ck,ck1},
\beq
\cl_\ps =
\half \psb (\ga^\mu i D_\mu - \m + \Qhat) \ps + {\rm h.c.} , 
\label{fermlag}
\eeq
where  
$\Qhat$ is a general $4\times4$ spinor matrix involving 
the covariant derivative $iD_\mu$ 
and the electromagnetic field tensor
$F_\ab \equiv \prt_\al A_\be - \prt_\be A_\al$.
The hermiticity of the Lagrange density \rf{fermlag} guarantees that 
$\Qhat$ satisfies condition $\Qhat = \ga_0 \Qhat^\dag\ga_0$.
In the limit of the free Dirac fermion with $A_\al = 0$,
Ref.~\cite{km13} has studied the propagation of the fermion field $\ps$
at arbitrary mass dimension.
A similar analysis of the quadratic terms in the photon sector 
at arbitrary mass dimension has been presented in \Ref{km09},
as well as extensions to other sectors,
e.g.,
nonminimal neutrino 
\cite{km12}
and gravity
\cite{nonmingrav}.

In this work we focus on the dominant Lorentz-violating effects 
including the photon-fermion interaction beyond the minimal SME
and restrict our attention to operators in the Lagrange density~\rf{fermlag}
with mass dimensions up to six.
The related full Lagrange density \rf{fermlag} can then be expressed
as two parts,
the conventional Lagrange density $\cl_0$
plus a series of contributions $\cl^{(d)}$ 
according to the mass dimension of the operators,
presented in Ref.~\cite{16dk}.
We note that the nonminimal operators in the Lagrange density~\rf{fermlag}
generate a new type of SME coefficients with subscript $F$
which control the sizes of interactions involving the fermion spinors $\ps$ 
and the electromagnetic field strength $F_\ab$.
For example, the dimension-five terms in the Lagrange density~\rf{fermlag} 
contain a contribution involving $\bcf 5 \mab F_\ab \psb \ga_5 \ga_\mu \ps$.

In the Lagrange density \rf{fermlag},
the presence of Lorentz-violating operators 
modifies the conventional Dirac equation for a fermion in electromagnetic fields
and generates corrections $\de\cH$ to the hamiltonian.
Since no Lorentz violation has been observed so far,
any corrections must be tiny.
We thus can treat $\de\cH$ as a perturbative contribution 
and apply perturbation theory to 
obtain the dominant Lorentz- and CPT-violating shifts in energy levels,
\beq
\de E_{n,s}= \vev{\ch_{n,s}|\de\cH|\ch_{n,s}},
\label{deltae}
\eeq
where $E_{n,s}$ are unperturbed eigenstates of $n$th level 
and $s$ is the spin state taking values of $+1$ and $-1$ for spin up and down,
respectively.

From the modified Dirac equation given by the Lagrange density \rf{fermlag}
\beq
(p \cdot \ga - m + \Qhat) \ps = 0,
\eeq
the exact hamiltonian $\cH$ can be defined as 
\beq
\cH \ps \equiv p^0 \ps = \ga_0 (\pvec \cdot \gavec + m - \Qhat)\ps,
\eeq
where $p^0$ is the exact energy. 
The exact perturbative hamiltonian $\de\cH$ can then be identified as
\beq
\de\cH = -\ga_0 \Qhat.
\eeq
It is challenging to construct $\de\cH$ directly 
as terms proportional to higher powers of momentum appear in $\Qhat$
and these terms in general contain the perturbative hamiltonian $\cH$ itself.
However, 
any contributions to $\de\cH$ due to the exact hamiltonian $\cH$ 
are at second or higher orders in the SME coefficients.
To obtain the leading-order corrections,
one thus can evaluate $p^0$ in $\Qhat$ at the unperturbed eigenstates $E_{n,s}$,
\beq
\de\cH \approx -\ga_0 \Qhat |_{p^0\to E_{n,s}}.
\eeq

In a Penning-trap experiment
the primary observables of interest are transition frequencies 
generated by the energy shifts due to the electromagnetic fields in the trap.
Among the key frequencies are
the Larmor frequency for spin precession $\nu_L \equiv \om_L/2\pi$
and the cyclotron frequency $\nu_c \equiv \om_c/2\pi$.
The difference of the two frequencies
gives the anomaly frequency $\nu_L-\nu_c = \nu_a \equiv  \om_a/2\pi$
\cite{geo}.
The measurements of the magnetic moment and the related $g$ factor of a particle 
confined in the trap can then be determined by the following ratio,
\beq
\fr {\nu_L}{\nu_c} \equiv \fr {\om_L}{\om_c} = \fr {g}{2}.
\label{ratio}
\eeq
The above frequencies can be shifted in the presence of Lorentz and CPT violation, 
as the energies are modified by \eq{deltae}.
To show the explicit results of the shifts,
we choose the apparatus frame with cartesian coordinates 
$x^a \equiv (x^1 , x^2, x^3)$
so that the magnetic field $\mbf B=B \hat x_3$
points at the positive $x^3$ direction 
and fix the electromagnetic potential gauge to be $A^\mu=(0, x_2B, 0,0)$.
For a confined particle of fermion-flavor $w=e$, $p$ for electrons and protons,
and of charge polarity $\si=+1$,~$-1$ for carrying positive and negative charges, 
there is no leading-order contribution from Lorentz and CPT violation 
to the cyclotron frequencies,
\beq
\de \om_c^{w} =
\de E_{1,\Z}^{w}-\de E_{0,\Z}^{w} 
\approx 0.
\label{cyclnochange}
\eeq
The dominant Lorentz- and CPT-violating contributions appear in 
the shifts to the anomaly frequencies,
\beq
\de \om_a^{w} 
=
\de E_{0,-\si}^{w}-\de E_{1,\si}^{w}
= 2 \btw w 3 - 2 \bftw w {33} B ,
\label{delomaw}
\eeq
where the tilde quantities are defined by
\bea
\btw w 3
&=&
\bcmw 3 3 w
+ \Hcmw 3 {12} w
- \mw \dcmw 4 {30} w
- \mw \gcmw 4 {120} w
+ \mw^2 \bcw 5 {300} w
+ \mw^2 \Hcw 5 {1200} w
- \mw^3 \dcw 6 {3000} w
- \mw^3 \gcw 6 {12000} w,
\nn\\
\bftw w {33}
&=&
\bcfw 5 {312} w
+ \Hcfw 5 {1212} w
- \mw \dcfw 6 {3012} w
- \mw \gcfw 6 {12012} w .
\label{tild}
\eea
Here the superscripts~$(d)$ of the nonminimal SME coefficients 
in the tilde quantities \rf{tild} denote the mass dimensions of the corresponding coefficients.

For the Lorentz- and CPT-violating shifts 
to the cyclotron and anomaly frequencies 
of the corresponding antifermion of flavor $\ol{w}$,
a similar analysis can be carried out 
by reversing the signs of the CPT-odd SME coefficients
in \eqs{cyclnochange} and \rf{delomaw}.
Similar to the fermion case,
the leading-order contributions to the cyclotron frequencies vanish,
\bea
\de \om_c^{\ol w} 
=
\de E_{1,\Z}^{\ol w}-\de E_{0,\Z}^{\ol w} 
\approx 0 ,
\eea
and the shifts to the anomaly frequencies are given by
\bea
\de \om_a^{\ol w}
=
\de E_{0,-\si}^{\ol w}-\de E_{1,\si}^{\ol w} 
= - 2 \btws w 3 + 2 \bftws w {33} B ,
\label{delomaws}
\eea
where the two sets of starred tilde coefficients are defined as 
\bea
\btws w 3
&=&
\bcmw 3 3 w
- \Hcmw 3 {12} w
+ \mw \dcmw 4 {30} w
- \mw \gcmw 4 {120} w
+ \mw^2 \bcw 5 {300} w
- \mw^2 \Hcw 5 {1200} w
+ \mw^3 \dcw 6 {3000} w
- \mw^3 \gcw 6 {12000} w,
\nn\\
\bftws w {33}
&=&
\bcfw 5 {312} w
- \Hcfw 5 {1212} w
+ \mw \dcfw 6 {3012} w
- \mw \gcfw 6 {12012} w .
\label{tilds}
\eea
The index pair 12 in the tilde quantities \rf{tild} and \rf{tilds} is antisymmetric 
and transforms under rotation like a single 3 index,
thus the shifts \rf{delomaw} and \rf{delomaws} in the anomaly frequencies 
for both fermions and antifermions depend on only the $\hat x_3$ direction,
as expected from the cylindrical symmetry of the trap.

The above results \rf{delomaw} and \rf{delomaws} 
show that the dominant contributions in the anomaly frequencies 
for a trapped fermion and antifermion of flavor $w$ in a Penning trap 
are given by the four tilde combinations
$\btw w 3$, $\bftw w {33}$, $\btws w 3$, and $\bftws w {33}$.
The results are valid in the apparatus frame,
in which the magnetic field is aligned with the positive $\hat x_3$ axis.
However, 
this apparatus frame is noninertial due to the Earth's rotation.
The standard canonical frame adopted in the literature to compare results 
from different experiments searching for Lorentz violation
is the Sun-centered frame 
\cite{sunframe0,sunframe},
with the cartesian coordinates $X^J\equiv (X,Y,Z)$.
In this frame,
the $Z$ axis is aligned with the rotation axis of the Earth
and the $X$ axis points towards the vernal equinox in the year~2000.
The Sun-centered frame is approximately inertial in a typical time scale for an experiment.
To relate the SME coefficients from the Sun-centered frame to the apparatus frame,
we introduce a third frame called the standard laboratory frame
with cartesian coordinates $x^j \equiv (x,y,z)$.
The $z$ axis in this frame points towards the local zenith and
the $x$ axis is aligned with the local south.
The choice of the positive $\hat x_3$ axis in the apparatus frame
to be aligned with the direction of the magnetic field 
may result in a nonzero angle to the $\hat z$ axis,
so the transformation $x^a = R^{aj} x^j$ 
relating $(x,y,z)$ in the standard laboratory frame  
to $(x^1,x^2,x^3)$ in the apparatus frame
involves a rotation matrix $R^{aj}$ specified in general by suitable Euler angels 
$\al$, $\be$, and $\ga$,
\bea
R^{aj} =
\left(
\begin{array}{ccc}
\cos\ga &\sin\ga  &0\\
-\sin\ga &\cos\ga &0\\
0 &0 &1\\
\end{array} 
\right)
\left(
\begin{array}{ccc}
\cos\be &0 &-\sin\be\\
0 &1 &0\\
\sin\be &0 &\cos\be\\
\end{array} 
\right)
\left(
\begin{array}{ccc}
\cos\al &\sin\al &0\\
-\sin\al &\cos\al &0\\
0 &0 &1\\
\end{array} 
\right).
\label{euler}
\eea
Neglecting boost effects, 
which are at the order of $10^{-4}$,
the relationship 
$x^{j} = R^{jJ} x^J$
between $(X, Y, Z)$ in the Sun-centered frame
and $(x,y,z)$ in the standard laboratory frame
can be obtained by applying the following rotation matrix
\cite{sunframe0,sunframe}
\beq
R^{jJ}(T_\oplus)=\left(
\begin{array}{ccc}
\cos\ch\cos\om_\oplus T_\oplus
&
\cos\ch\sin\om_\oplus T_\oplus
&
-\sin\ch
\\
-\sin\om_\oplus T_\oplus
&
\cos\om_\oplus T_\oplus
&
0
\\
\sin\ch\cos\om_\oplus T_\oplus
&
\sin\ch\sin\om_\oplus T_\oplus
&
\cos\ch
\end{array}
\right),
\label{rotmat}
\eeq
where $\om_\oplus\simeq 2\pi/(23{\rm ~h} ~56{\rm ~min})$
is the sidereal frequency of the Earth's rotation,
$T_\oplus$ is the local sidereal time,
and the angle $\ch$ specifies the laboratory colatitude.

To relate the time $T$ in the Sun-centered frame  
to the time $t$ in the standard laboratory frame,
it is often convenient to match the origin of $t$
with the local sidereal time $T_\oplus$,
by defining its origin at the moment 
when the $y$ axis in the standard laboratory frame 
lies along the $Y$ axis in the Sun-centered frame.
For a laboratory with longitude $\la$ in units of degrees,
this choice offsets $t$ from $T$ by
an integer number of the Earth's sidereal rotations
plus an additional shift 
\beq
T_0 \equiv T-T_\oplus 
\simeq 
\fr{(66.25^\circ- \la)}{360^\circ}
(23.934~{\rm hr}).
\label{T0}
\eeq

The above discussion shows that the relationship between
$(X,Y,Z)$ in the Sun-centered frame
and $(x^1,x^2,x^3)$ in the apparatus frame is given by
\beq
x^{a} (T_\oplus) 
= R^{a j} R^{jJ} (T_\oplus) X^J.
\label{transf}
\eeq
The transformation \rf{transf} generates the dependence on the sidereal time
of the SME coefficients observed in the apparatus frame.
To show the explicit dependence of the shifts to the anomaly frequencies \rf{delomaw},
consider a fermion of flavor $w$ confined in a Penning trap 
with the magnetic field aligned with the local zenith and located at colatitude $\ch$,
applying the transformation matrix \rf{rotmat} yields the results
\beq
\btw w 3 =
\btw w Z \cos\ch 
+ ( \btw w X \cos\om_\oplus T_\oplus 
+ \btw w Y \sin\om_\oplus T_\oplus )\sin\ch ,
\eeq
and 
\bea
\bftw w {33}
&=&
\bftw w {ZZ}
+\half (\bftw w {XX} +\bftw w {YY} -2\bftw w {ZZ}) \sin^2\ch
+( \bftw w {(XZ)} \cos\om_\oplus T_\oplus
+ \bftw w {(YZ)} \sin\om_\oplus T_\oplus )\sin 2\chi 
\nn\\ 
&&
+\Big(
\half (\bftw w {XX} - \bftw w {YY}) \cos 2\om_\oplus T_\oplus
+ \bftw w {(XY)} \sin 2\om_\oplus T_\oplus \Big) \sin^2\ch ,
\eea
where the parenthesis around two indices $(JK)$ 
in the tilde coefficients means symmetrization
and is defined as $(JK)=(JK+KJ)/2$.
Similar results for the shifts to the anomaly frequencies \rf{delomaws} 
of antifermions can also be derived  
by substituting the tilde coefficients with the starred tilde coefficients.
In more general cases with the magnetic field pointing a generic direction,
information on the Euler angles $\al$, $\be$, $\ga$ in \eq{euler} 
are needed to obtain the explicit results.

The results given above show that the physical observables 
in a Penning-trap experiment involving 
electrons, positrons, protons, and antiprotons 
are the 36 independent tilde quantities
$\btw w J$, $\btws w J$, $\bftw w {(JK)}$, and $\bftws w {(JK)}$
in the Sun-centered frame.
Performing a sidereal-variation analysis of the anomaly frequencies can give access to
28 of the coefficients as the other 8 contributions proportional to
$\btw w Z$, $\btws w Z$, $\bftw w {ZZ}$, and $\bftws w {(ZZ)}$
are independent of sidereal time. 
A comparison of the results from two different Penning-trap experiments 
is therefore required to study these 8 combinations of coefficients for Lorentz violation 
that produce constant shifts to the anomaly frequencies.

\section{Experiment}
\label{experiment}

\subsection{Harvard Experiment}

The recent measurement of the electron $g$ factor 
performed at Harvard University has reached a precision of 0.28 ppt
\cite{ha11}.
A sidereal-variation analysis of the anomaly frequencies for the electron
was performed to search for variations in the sidereal time of the Earth's rotation
\cite{07ha}.
The data was analyzed for oscillations over time 
and was fit by a five-parameter sinusoid model
at the sidereal frequencies of $\om_{\oplus}$ and $2\om_{\oplus}$,
yielding a 2$\si$ limit in the amplitudes of the harmonic oscillation of
$|\de \nu_a^e| \lsim 0.05  {\rm\ Hz}$.
This result corresponds to $|\de \om_a^e| \lsim 2\times 10^{-25}$ GeV 
in natural units with $c = \hbar = 1$.
The magnetic field adopted in the experiment is $B=5.36$ T in the local upward direction
and the geometrical colatitude of this experiment is $\ch=47.6^\circ$.
Taking one sidereal oscillation at a time places bounds 
\bea
&&
\hskip-10pt
\bigg(\Big( \btw e X - (2 \times 10^{-15}{\rm ~GeV}^2)\bftw e {(XZ)}\Big)^2 
+ \Big( \btw e Y - (2 \times 10^{-15}{\rm ~GeV}^2)\bftw e {(YZ)}\Big)^2
\bigg)^{1/2} 
\lsim 2\times 10^{-25} {\rm ~GeV}
\label{harvard1st}
\eea
in the first harmonic 
and 
\bea
&&
\hskip-10pt
\bigg(\Big(10^{-15} {\rm ~GeV}^2(\bftw e {XX} - \bftw e {YY})\Big)^2 
+ \Big(10^{-15} {\rm ~GeV}^2 \bftw e {(XY)}\Big)^2 
\bigg)^{1/2} 
\lsim 2\times  10^{-25} {\rm ~GeV}
\label{harvard2rd}
\eea
in the second harmonic,
respectively.
The above results not only lead to a factor of four improvement
compared to the existing constraints obtained by a similar analysis
of the Penning-trap experiment searching for first-harmonic variation
at the University of Washington
\cite{mi99, 16dk},
but also produce the first-time bounds on tilde coefficients 
$\bftw e {(XX)}-\bftw e {(YY)}$ and $\bftw e {(XY)}$ 
as they only appear in the second harmonic of the sidereal oscillation.

The experiments to measure the magnetic moment of a trapped positron
are currently under development at Harvard University and Northwestern University
\cite{15fo, 19gf}.
Performing a similar sidereal-variation analysis of the anomaly frequency 
would offer not only the first-time limits
on the starred tilde coefficients $\btws e J$, $\bftws e {(JK)}$,
but would also constrain the CPT-odd coefficients 
in \eqs{tild} and \rf{tilds}
by comparing with measurements of the electron.
The constant parts in the sidereal variations of the tilde coefficients
$\btw e J$, $\bftw e {(JK)}$, $\btws e J$, and $\bftws e {(JK)}$
could also be studied by this comparison.

\subsection{BASE Experiments at Mainz and CERN}

The BASE collaboration has recently measured the proton magnetic moment 
at a record sensitivity of 0.3 ppb using a Penning trap located at Mainz
\cite{17sc},
improving their previous best result 
\cite{mo14}
by a factor of 11.
A precision of 1.5 ppb of the antiproton magnetic moment measurement
has also been acheived by the same group 
using a similar Penning trap located at CERN
\cite{17sm}.
A study of sidereal variations of the anomaly frequencies 
for both protons and antiprotons is currently being performed at BASE and this could, 
in principle, 
provide sensitivities to various tilde coefficients 
$\btw p J$, $\bftw p {(JK)}$, $\btws p J$ and $\bftws p {(JK)}$.
Another version of this experiment is planned to be performed 
at CERN by the BASE collaboration 
to measure the magnetic moments
for both protons and antiprotons using quantum logic readout 
\cite{chip},
which will allow rapid measurements of the anomaly frequencies for the proton and antiproton.
This would offer an excellent opportunity to conduct the sidereal-variation analysis,
as well as to constrain the constant parts in the harmonics of the above coefficients
through a direct comparison of the two measurements.

Here we combine the published results from the two recent BASE experiments
\cite{17sc, 17sm}
to obtain constraints on the SME coefficients in the Sun-centered frame.
A comparison between the two measured $g$ factors for protons and antiprotons gives
\bea
\fr{g_p}{2} - \fr{g_{\ol{p}}}{2}
=
\fr {\om_a^p}{\om_c^p} - \fr {\om_a^{\ol p}}{\om_c^{\ol p}} 
=
\fr 2 {\om_c^p \om_c^{\ol p}} 
\left( \Si \om_c^{p} \De \om_a^{p} 
- \De \om_c^{p} \Si \om_a^{p}\right),
\label{sumdiff}
\eea
where the differences and sums of the cyclotron and anomaly frequencies are
defined as
\bea
\De \om_c^{p} &\equiv&
\half (\om_c^{p} - \om_c^{\ol p}),
\nn\\
\Si \om_c^{p} &\equiv&
\half (\om_c^{p} + \om_c^{\ol p}),
\nn\\
\De \om_a^{p} &\equiv&
\half (\de \om_a^{p} - \de \om_a^{\ol p}),
\nn\\
\Si \om_a^{p} &\equiv&
\half (\de \om_a^{p} + \de \om_a^{\ol p}).
\eea
For the proton magnetic moment measured at Mainz,
the experiment is located at $\ch \simeq 40.0^\circ$ 
and the applied magnetic field $B\simeq 1.9$ T points 
$\th=18^\circ$ from local south in the counterclockwise direction,
generating a cyclotron frequency $\om_c^{p}=2\pi\times 28.96$ MHz
\cite{17sc}.
For the antiproton magnetic moment measurement at CERN,
the trap is located at $\ch^* \simeq 43.8^\circ$ 
and the magnetic field $B^*\simeq 1.95$ T points 
$\th^*=120^\circ$ from local south in the counterclockwise direction,
producing a different cyclotron frequency 
$\om_c^{\ol p}=2\pi\times 29.66$ MHz
\cite{17sm}.
Since the measurements of the frequencies for both experiments 
were performed over an extended time period,
any sidereal variations could be plausibly assumed to be averaged out,
leaving only the constant parts in the tilde coefficients.
Therefore,
applying the general transformation \rf{transf} together with  
the related experimental quantities 
yields the following expressions for 
the time-independent parts in $\De \om_a^{p}$ and $\Si \om_a^{p}$,
\bea
\De \om_a^{p} 
&=& 
\btw p 3 - \bftw p {33} B 
+ \btws p 3 - \bftws p {33} B^*
\nn\\
&=& 
- \btw p Z \cos\th\sin\ch
- \btws p Z \cos\th^*\sin\ch^*
\nn\\
&&
- \half (\bftw p {XX} + \bftw p {YY}) B (\cos^2\th \cos^2\ch+\sin^2\th) 
- \half (\bftws p {XX} + \bftws p {YY}) B^* (\cos^2\th^* \cos^2\ch^*+\sin^2\th^*) 
\nn\\
&&
- \bftw p {ZZ} B \cos^2\th \sin^2\ch 
- \bftws p {ZZ} B^* \cos^2\th^* \sin^2\ch^*  ,
\nn\\
\Si \om_a^{p} 
&=& 
\btw p 3 - \bftw p {33} B 
- \btws p 3 + \bftws p {33} B^*
\nn\\
&=& 
- \btw p Z \cos\th\sin\ch
+ \btws p Z \cos\th^*\sin\ch^*
\nn\\
&&
- \half (\bftw p {XX} + \bftw p {YY}) B (\cos^2\th \cos^2\ch+\sin^2\th) 
+ \half (\bftws p {XX} + \bftws p {YY}) B^* (\cos^2\th^* \cos^2\ch^*+\sin^2\th^*) 
\nn\\
&&
- \bftw p {ZZ} B \cos^2\th \sin^2\ch 
+ \bftws p {ZZ} B^* \cos^2\th^* \sin^2\ch^* .
\label{delsum}
\eea
Substituting expressions \rf{delsum} into the difference \rf{sumdiff}
and adopting the numerical values of the experimental quantities given above,
the reported results for the measurements of $g$ factors 
from both BASE experiments give the following 2$\si$ limit
\bea
&&\Big| \btw p Z - 0.6 \btws p Z 
+ (2\times 10^{-16}{\rm~ GeV}^{2})(\bftw p {XX} + \bftw p {YY}) 
+ (2\times 10^{-16}{\rm~ GeV}^{2})\bftw p {ZZ} 
\nn\\
&&
\hskip 10pt
+ (2\times 10^{-16}{\rm~ GeV}^{2})(\bftws p {XX} + \bftws p {YY}) 
+ (7\times 10^{-17}{\rm~ GeV}^{2})\bftws p {ZZ} 
\Big|
\lsim 8\times 10^{-25} {\rm ~GeV}.
\qquad
\label{basecomp}
\eea

\section{Sensitivity} 
\label{sensitivity}

To get some intuition for the scope of the constraints 
\rf{harvard1st}, \rf{harvard2rd}, and \rf{basecomp},
a common practice is to assume only one individual tilde coefficient is nonzero at a time.
Considering no Lorentz and CPT violation has been observed so far,
this procedure offers a reasonable measure of the estimated limits
on each tilde coefficient by ignoring any cancellations among them.
We list in Table~\ref{epconstraints} 
the resulting constraints on the tilde coefficients from this work
and also include the previous limits obtained in \Ref{16dk},
as well as recent improved results presented in \Ref{17sm},
for a direct comparison. 
In the electron sector,
Table~\ref{epconstraints} shows that not only a factor of four improvement 
for the limits on the tilde coefficients 
$\btw e X$, $\btw e Y$, $\btw e {(XZ)}$, and $\btw e {(YZ)}$
has been achieved,
but also that new coefficients 
$\bftw e {(XY)}$ and $\bftws e {XX} - \bftws e {YY}$
have been constrained.
In the proton sector, 
the limits on the tilde coefficients have been improved
by factors of up to three compared to the existing results
\cite{17sm}.
The constraints on the tilde coefficients that are not sensitive 
to the corresponding work are left blank in Table~\ref{epconstraints}.
Note that only 18 out of the 36 coefficients for Lorentz violation 
related to Penning-trap experiments have been constrained so far.
A sidereal-variation analysis for the measurements of the magnetic moments
of protons and antiprotons would permit access to other various components 
of the tilde coefficients in the proton sector.

\renewcommand{\arraystretch}{1.5}
\begin{table*}
	\centering
	\caption{
		\label{epconstraints}
		New and improved constraints on the SME coefficients.}
	\setlength{\tabcolsep}{5pt}
	\begin{tabular}{clll}
		\hline
		\hline															
		Coefficient			&			Previous Constraint in \cite{16dk}			&			Recent Result in \cite{17sm}			&			This Work			\\	\hline
$	|	\btw e X	|	$	&	$	<	6\times 10^{-25}	{\rm ~GeV}	$	&	$				$	&	$	<	1\times 10^{-25}	{\rm ~GeV}	$	\\	
$	|	\btw e Y	|	$	&	$	<	6\times 10^{-25}	{\rm ~GeV}	$	&	$				$	&	$	<	1\times 10^{-25}	{\rm ~GeV}	$	\\	
$	|	\btw e Z	|	$	&	$	<	7\times 10^{-24}	{\rm ~GeV}	$	&	$				$	&	$				$	\\	
$	|	\btws e Z	|	$	&	$	<	7\times 10^{-24}	{\rm ~GeV}	$	&	$				$	&	$				$	\\	
$	|	\bftw e {XX} + \bftw e {YY}	|	$	&	$	<	2\times 10^{-8}	{\rm ~GeV}^{-1}	$	&	$				$	&	$				$	\\	
$	|	\bftw e {ZZ}	|	$	&	$	<	8\times 10^{-9}	{\rm ~GeV}^{-1}	$	&	$				$	&	$				$	\\	
$	|	\bftw e {(XY)} 	|	$	&	$				$	&	$				$	&	$	<	2\times 10^{-10}	{\rm ~GeV}^{-1}	$	\\	
$	|	\bftw e {(XZ)} 	|	$	&	$	<	4\times 10^{-10}	{\rm ~GeV}^{-1}	$	&	$				$	&	$	<	1\times 10^{-10}	{\rm ~GeV}^{-1}	$	\\	
$	|	\bftw e {(YZ)} 	|	$	&	$	<	4\times 10^{-10}	{\rm ~GeV}^{-1}	$	&	$				$	&	$	<	1\times 10^{-10}	{\rm ~GeV}^{-1}	$	\\	
$	|	\bftws e {XX} + \bftws e {YY}	|	$	&	$	<	2\times 10^{-8}	{\rm ~GeV}^{-1}	$	&	$				$	&	$				$	\\	
$	|	\bftws e {XX} - \bftws e {YY}	|	$	&	$				$	&	$				$	&	$	<	4\times 10^{-10}	{\rm ~GeV}^{-1}	$	\\	
$	|	\bftws e {ZZ}	|	$	&	$	<	8\times 10^{-9}	{\rm ~GeV}^{-1}	$	&	$				$	&	$				$	\\	\hline
$	|	\btw p Z	|	$	&	$	<	2\times 10^{-21}	{\rm ~GeV}	$	&	$	<	1.8\times 10^{-24}	{\rm ~GeV}	$	&	$	<	8\times 10^{-25}	{\rm ~GeV}	$	\\	
$	|	\btws p Z	|	$	&	$	<	6\times 10^{-21}	{\rm ~GeV}	$	&	$	<	3.5\times 10^{-24}	{\rm ~GeV}	$	&	$	<	1\times 10^{-24}	{\rm ~GeV}	$	\\	
$	|	\bftw p {XX} + \bftw p {YY}	|	$	&	$	<	1\times 10^{-5}	{\rm ~GeV}^{-1}	$	&	$	<	1.1\times 10^{-8}	{\rm ~GeV}^{-1}	$	&	$	<	4\times 10^{-9}	{\rm ~GeV}^{-1}	$	\\	
$	|	\bftw p {ZZ}	|	$	&	$	<	1\times 10^{-5}	{\rm ~GeV}^{-1}	$	&	$	<	7.8\times 10^{-9}	{\rm ~GeV}^{-1}	$	&	$	<	3\times 10^{-9}	{\rm ~GeV}^{-1}	$	\\	
$	|	\bftws p {XX} + \bftws p {YY}	|	$	&	$	<	2\times 10^{-5}	{\rm ~GeV}^{-1}	$	&	$	<	7.4\times 10^{-9}	{\rm ~GeV}^{-1}	$	&	$	<	3\times 10^{-9}	{\rm ~GeV}^{-1}	$	\\	
$	|	\bftws p {ZZ}	|	$	&	$	<	8\times 10^{-6}	{\rm ~GeV}^{-1}	$	&	$	<	2.7\times 10^{-8}	{\rm ~GeV}^{-1}	$	&	$	<	1\times 10^{-8}	{\rm ~GeV}^{-1}	$	\\						\hline
		\hline
	\end{tabular}
\end{table*}

\section{Summary} 

In conclusion,
we present in this work the general theory 
for quantum electrodynamics
with Lorentz- and CPT-violating operators of mass dimensions up to six
and study the dominant effects arising from Lorentz and CPT violation
in Penning-trap experiments involving confined particles.
Recently reported results of magnetic moments of the confined particles
are used to improve existing bounds 
on various SME coefficients,
and to constrain new coefficients as well.
The results obtained in this work are summarized in Table~\ref{epconstraints}.
The methodology we outline in this work using \eq{sumdiff} to derive these constraints 
can be used as a generic way to study Lorentz and CPT violation
involving comparisons of results from different Penning-trap experiments.
The high sensitivities of the measurements 
in current and forthcoming experiments offer
strong motivation to continue the efforts of studying 
Lorentz and CPT violation with great potential 
to uncover any possible tiny signals.

\vspace{8pt} 

\acknowledgments{
The author would like to thank Jay Tasson for the invitation and 
V. Alan~Kosteleck\'y for useful discussion.
This research was funded in part by the Department of Energy grant number {DE}-SC0010120
and by the Indiana University Center for Spacetime Symmetries.} 

\conflictsofinterest{The author declares no conflict of interest.} 

\appendixtitles{yes} 
\appendixsections{one} 



\reftitle{References}



\end{document}